# Vortex interaction with a rough wall formed by a hexagonal lattice of posts

Qianhui Li and Christoph H. Bruecker

School of Mathematics, Computer Science and Engineering. City, University of London, UK

An experimental study is reported which investigates the head-on collision of a laminar vortex ring of diameter D ($Re_\Gamma = 3000$) on a fakir-like surface composed of circular posts of height h/D=0.068 located on a planar bed. Lattices of the posts in hexagonal and random distribution (average porosity of ϵ=0.94 in the layer) are compared to each other with respect to the plain wall. Prior to impact, the vortex ring develops the early state of natural azimuthal instabilities of different mode numbers N=5-7 competing with each other. While impacting with the rough wall, a secondary ring is observed which is pushed outwards and is not wrapped around the primary ring as in flat wall impact. Between both rings of opposite sign vorticity, a strong fluid rebound is induced. The hexagonal lattice causes the rapid growth of further secondary vortex structures in a regular mode number *N*=6 arrangement at the outer edge of the primary ring in form of six lobes which are aligned with the orientations of preferential "pathways" in the layer. At the outer tip of the lobes radial wall-jets are generated. Rotating the fakir geometry around the centre of impact also rotates the jets' location and direction accordingly. A surface with random lattice of the posts at the same average number density is not able to repeat this observation and no regular secondary flow pattern is visible until full breakdown of the ring. The results show that a tailored arrangement of such posts can be used for near-wall flow control when patterns of preferred pathways in the posts' layer lock-on with existing instability modes such as in impacting jet flows or in turbulent boundary layer flows.

## 1. Introduction

Rough walls have attracted recently new attention in fluid mechanics. Surprising results were achieved to show transition delay with certain roughness while others showed promoting transition to turbulence (Fransson et al. 2006). This depends on the size and mechanism by which the roughness interacts with the near wall flow and the coherent vortices embedded in it (Rosti et al. 2018). Ring-type vortices have been understood as a reference structure to investigate the interaction of coherent vortex structures interacting with the wall. Therefore, model studies with vortex rings impacting on walls are a tool to investigate the near-wall cycle of self-replication of turbulent structures (Lim et al.

1991, Chu et al. 1995, Xu and Wang 2013). A prominent example in transitional and turbulent boundary layer flows is the structure of a hair-pin vortex impacting with the wall, inducing itself a train of trailing hairpin-type vortices which is called the wall cycle (Jimenez 1994, Jimenez and Pinelli 1999). Walker et al. (1987) studied experimentally a wide variety of laminar rings impacting with a wall in head-on collision, using dye in water to visualize the flow in the ring as well as near the plane surface. Their results show that an unsteady separation develops in the boundary-layer flow, in the form of a secondary ring attached to the wall. As a result, one vortex ring can produce another, through an unsteady interaction with a viscous flow near the wall. Studies on the aspect of rebound clearly revealed the origin of the boundary layer at the wall and the formation of the secondary ring as a source of this rebound (Orlandi & Verzicco 1993). However, not much is known about the influence of wall roughness on such behaviour. The only study reported so far on the interaction of a vortex ring with roughness is the study of a collision of a vortex ring with a porous screen (Adhikari & Lim 2009). This, however, differs largely from a rough wall at it allows the penetration of the vorticity through the screen.

In this work, we present results for a head-on collision of a laminar vortex ring with a rough wall made of a bed of slender posts. A set of flow realisations is obtained by modifying the preferred direction of permeability of the layer using different lattices of the posts. It will be shown that the texture triggers the transition of the vortex ring and increases the rebound while interacting with the wall. In particular, by tailoring the directionality of the preferred pathways one can lock-on the dominant mode and phase of evolving non-linearity in form of secondary vortices. This leads herein to the formation of radial wall-jets in the pattern of a ship wheel at the outer edge of the ring.

## 2. Experimental set-up

The experimental setup as shown in figure 1 follows that in the experiment of Ponitz et al. (2016) and it is briefly discussed here. The measurement is carried out in a liquid tank with octagonal cross-section. The present vortex ring is generated through a piston/cylinder nozzle installed at the top of the tank. The nozzle outlet has a diameter of 30 mm. The vortex ring has a radius of $R = D/2$ = 22 mm at an initial travelling speed of $U_0$ = 400 mm/s. The calculation of ring circulation follows $\Gamma_0 = \int \omega_\theta dr dz$ and is calculated to 180 cm²/s. This results in a circulation-based Reynolds number $Re_\Gamma = \Gamma_0/\nu$ = 3000. The time is made non-dimensional in the form $t^* = t\Gamma_0/R^2$ (it multiplies 38/s with the physical time in seconds) and starts at zero when the roll-up process of the shear layer at the nozzle is finished.

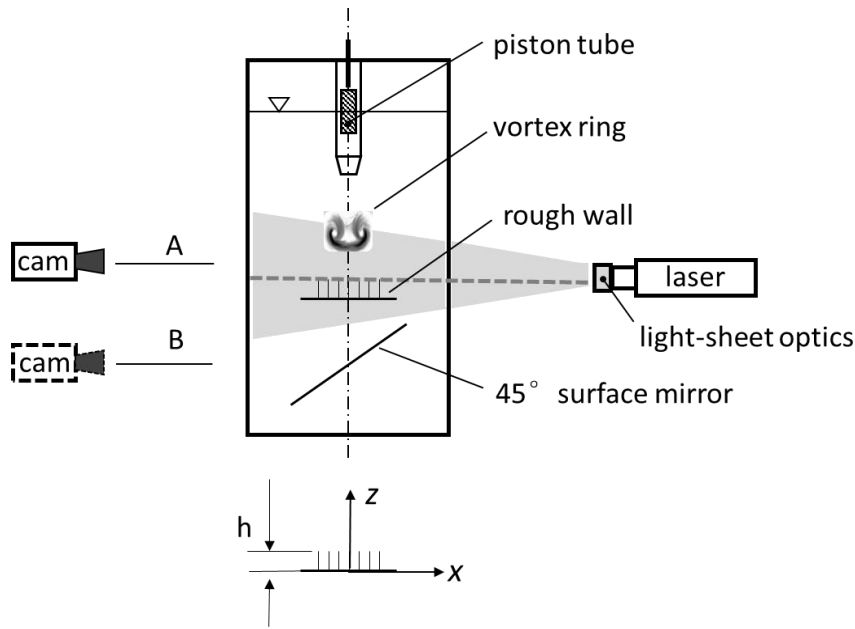

**Figure 1.** Schematic of the experimental setup. For Time-Resolved Particle Image Velocimetry (TR-PIV) measurements, the camera configuration A is used with a vertical light-sheet to record the flow field in the vertical *x-z* plane; configuration B is used to record the flow field in the radial *x-y* plane above of the interface. The coordinate origin is at the centre of the axis of the vortex ring on the surface of the wall.

Downstream from the nozzle exit at a distance of 20cm, a glass plate is placed inside the tank in the horizontal plane. As the vortex ring travels downstream, it impacts in head-on collision with the wall. In the experiments we compare the impact on a smooth glass wall with the one on the fakir geometry. The latter is made out of slender posts of height h which protrude out of a 2mm thick disc of diameter $D_{wall}$ = 110 mm. The material used to cast this fakir geometry is a transparent silicone (Poly-Di-methyl-Siloxane PDMS, Wacker Silicones). An aqueous sugar solution (44 Vol % glycerol to 56 Vol % water, density $\rho = 1.13\ g/cm^3$, kinematic viscosity $\nu = 6 \times 10^{-6} m^2/s$) is used as working liquid which matches the refractive index of the silicone material ($n = 1.4$). This allows optical access to the light-sheet plane trough the transparent surface. The cylindrical posts have a diameter of $d = 1mm$ and a height of $h$ = 3mm. Below the glass plate, a 45° titled surface mirror is arranged to allow optical access through the glass plate into the layer of the PIV plane. The number density of the cylindrical elements is 583 that are positioned on the disc with an average spacing of *Δs* = 5*d*. The resulting average porosity of this layer calculated by the ratio of void volume to total volume is $\epsilon = 0.94$. Note that variations in permeability in the wall-parallel plane or wall-normal direction can be designed by placing the posts in different arrangements.

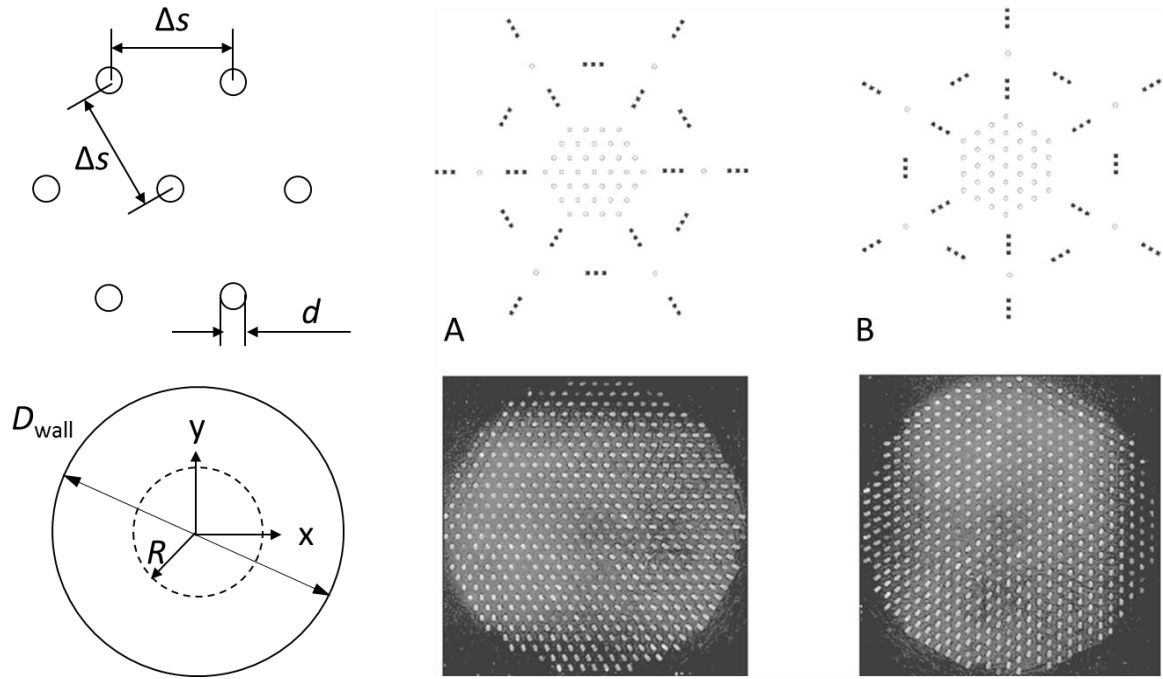

**Figure 2.** Arrangement of the fakir-like surface with posts on a circular disc of radius R in hexagonal lattice. The diameter of the elements is $d$ = 1mm and the spacing is $\Delta s$ = 5mm (the torus of the vortex ring with radius $R$ is indicated as a dashed circle). A random arrangement of the same number density of the elements is also *used* to compare with (not shown here). Pictures of the bed are taken from bottom with illumination using the horizontal laser sheet aligned along the tips. A: hexagonal lattice with rows of posts in horizontal direction. B: same as A but rotated about $\pi/2$.

The PIV imaging system is comprised of a high-speed camera (Phantom Miro 310/311, Ametek) with CMOS sensor of 1280×800 pixels recording at 2000 frames per second, equipped with a lens (Tokima Macro $f$ = 100mm, F 2.8). The imaging magnification factor is $M$ = 0.15. A continuous wave Argon-Ion laser (Coherent Innova 70C, 3 W power at $\lambda$ = 528 nm wavelength) is used as illumination source. The output laser beam is about 1.5 mm in diameter and is further expanded to a sheet. For the experiments the laser sheet is arranged in two variants: firstly, the vortex ring impact is in a horizontal plane 1mm above the interface layer in the fluid. Neutrally buoyant particles with a nominal diameter of 30 μm are chosen as flow tracer. The present measurement starts from the generation of the vortex ring and finishes when it undergoes final breakdown after impact. A total of 8000 particle images are acquired, corresponding to a measurement duration of 4 s.

The data post-processing contains image pre-processing, 2D cross-correlation of successive images to calculate the vectors following an iterative grid refinement method. The final interrogation window has a size of 32×32 pixels and processing is done on a grid with 50% overlap ratio. The resulting vector grid is then used to calculate the out-of-plane component of the vorticity vector. The finite-time Lyapunov exponent (FTLE) was then computed. Regions of high FTLE values allow to recognize the boundary between fluid contained in the vortex ring and fluid originating from the wall. For the case of the radial measurement plane the equidistant Cartesian velocity grid is later interpolated onto a polar-type grid with constant spacing in radial and azimuthal direction. A procedure to decompose the dominant mode n in azimuthal direction from measurement noise is to take the n-average in azimuthal direction for all vectors along a radial line at a given angle α, which is calculated as follows:

$$\left(\begin{array}{c}\overline{v_r}\\ \overline{v_\theta}\end{array}\right)\bigg|_\alpha = \frac{1}{n+1}\sum_{k=0}^{n}\begin{pmatrix}\cos\alpha_k & -\sin\alpha_k\\ \sin\alpha_k & \cos\alpha_k\end{pmatrix}\begin{pmatrix}v_r\\ v_\theta\end{pmatrix}\bigg|_{\alpha+\alpha_k}, \quad \alpha_k = k\,2\pi/n \quad \text{(eq. 1)}$$

This is done for the whole range of α in the polar vector diagram. The residual to the original field is:

$$\begin{pmatrix}v_r\\ v_\theta\end{pmatrix}\bigg|_{res,\alpha} = \begin{pmatrix}v_r\\ v_\theta\end{pmatrix} - \begin{pmatrix}\overline{v_r}\\ \overline{v_\theta}\end{pmatrix} \quad \text{(eq. 2)}$$

In a perfect polar-symmetric case with a single dominant azimuthal mode n, the averaging procedure in eq. (1) does not change the results and the residual is zero. It is used herein to highlight the dominant mode n field against noise if the residual difference vector field is of only low energy.

## 3. Results

Volumetric flow measurements of the vortex ring travelling freely in the tank were done in our lab in previous work (Sun and Bruecker 2017) to characterize the growth of azimuthal instabilities. The results for the axial vorticity $\omega_z$, which is representative of the transition process, show instabilities evolving in mode 5-7, while mode 6 is the most unstable wave predicted by the theory of Saffman (1978) for the studied vortex rings. The non-dimensional time of impact in the present study is defined at the moment when the free travelling vortex passes the origin of the coordinate system at $t_0^* = 15$. An additional time scale $T = t^* - t_0^*$ is defined to compare the different experiments after impact. At the moment of impact with the wall the vortex ring is within the early transition which is represented by similar amplitude of wavenumber *N*=5-7 along the core.

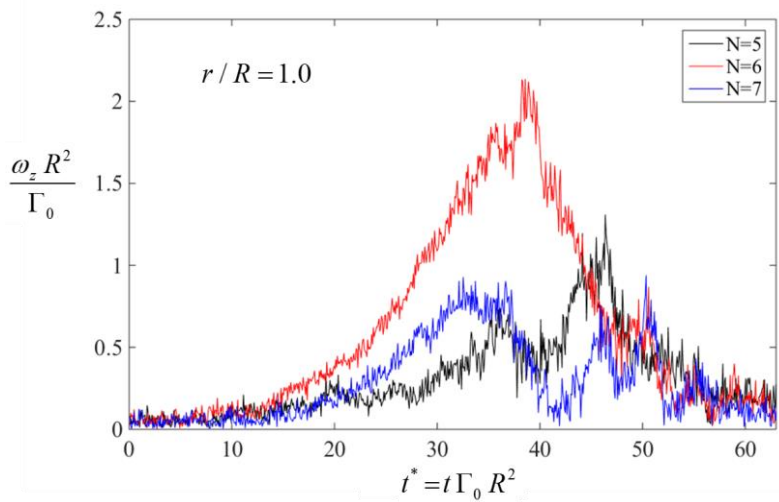

**Figure 3.** Evolution of amplitude of azimuthal wavenumber 5-7 in the axial vorticiy of the free travelling vortex ring, measured from Scanning Tomo-PIV in the same apparatus.

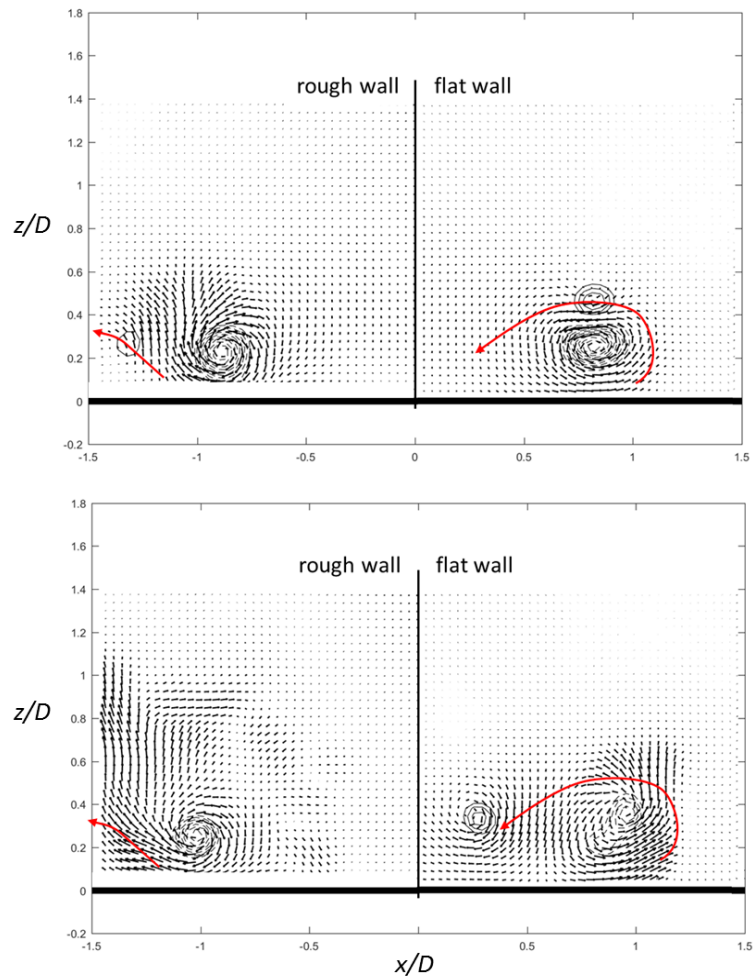

**Figure 4**: Velocity field snapshots in the vertical x-y plane for the rough and flat wall at *T*=2.71 and *T*=7. Iso-contours of $\omega_z$ indicate the location of the primary and secondary vortex ring (dashed line: primary vortex, solid line: secondary vortex). The red arrows indicate the passage of the secondary vortex over time.

Figure 4 displays the comparison of flow evolution for the reference case of the plane wall compared to the wall with the hexagonal lattice of posts. While the initial phase of approach and first contact with the bed surface is similar to the plane wall, one can see clear differences in later times. For the flat wall, the roll-up of the wall boundary layer into a secondary vortex ring is clearly visible. This secondary vortex ring interacts with the primary vortex ring by mutual induction. As the primary vortex ring expands to a radius of 0.9*D* it lifts the secondary ring up and wraps it around further towards the inner region. In comparison, for the fakir geometry the secondary ring is weaker and not being wrapped up by the primary ring, rather it is pushed radially outwards and remains in the same radial plane as the primary one. As both rings have opposite vorticity, a strong vertical rebound of the fluid is seen in the gap between the cores. In contrast, for the plain wall the fluid rebound starts later and is considerable weaker compared to the rough wall.

The flow penetrating into the posts layer is affected by the dissipation imposed by the posts and boundary layer separation is seemingly not coherent enough to roll-up the shear layer into a circular ring of concentrated vorticity. The difference to the plain wall is clearly seen by visualizing the FTL-exponent, see Fig. 5.

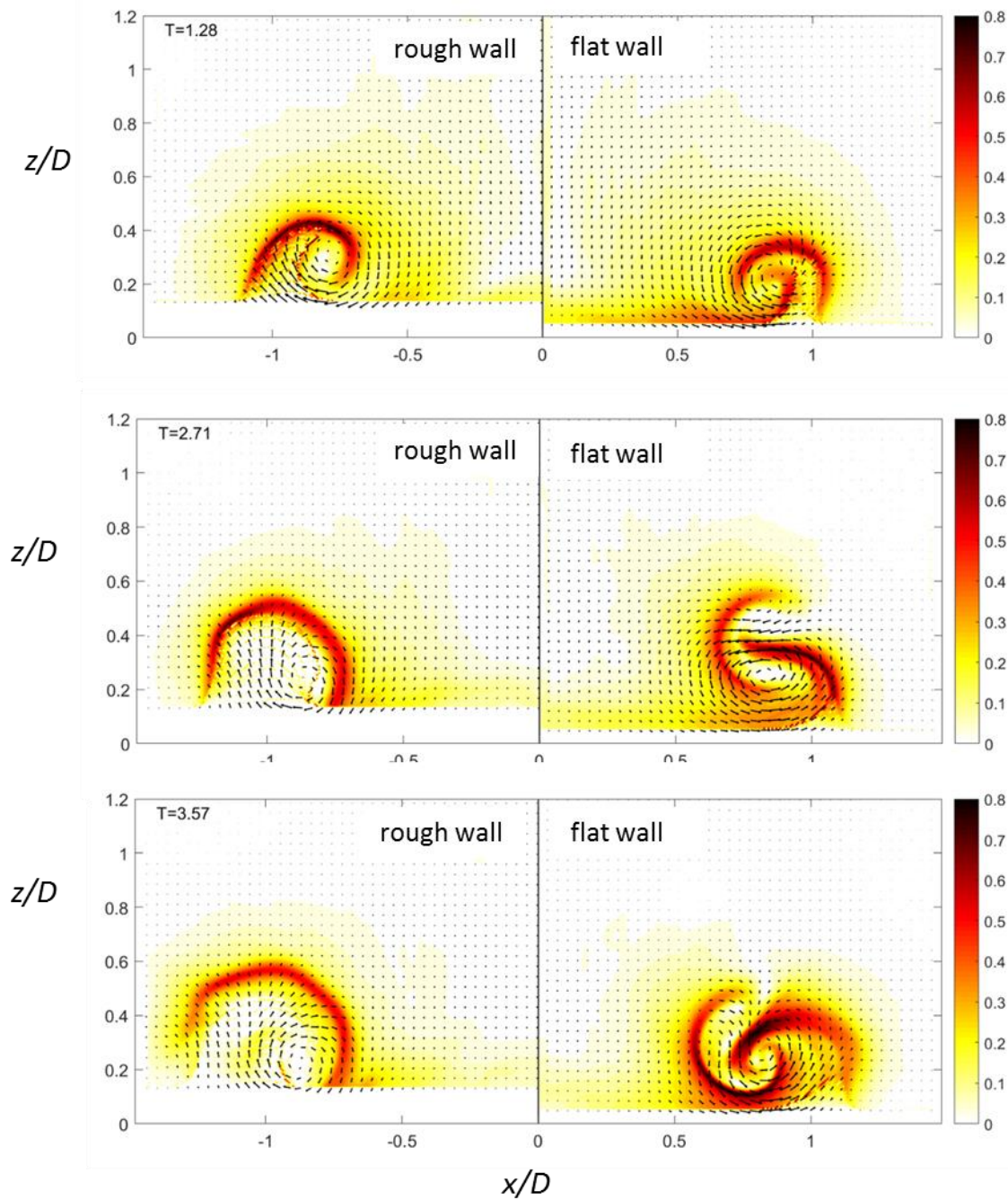

**Figure 5.** Evolution of the vortex roll-up in the vertical x-y plane visualized by means of the FLT-exponent in color. Left: fakir-geometry with hexagonal lattice, right: flat wall.

Further information is gained from the flow patterns in the radial plane above the tips of the posts.

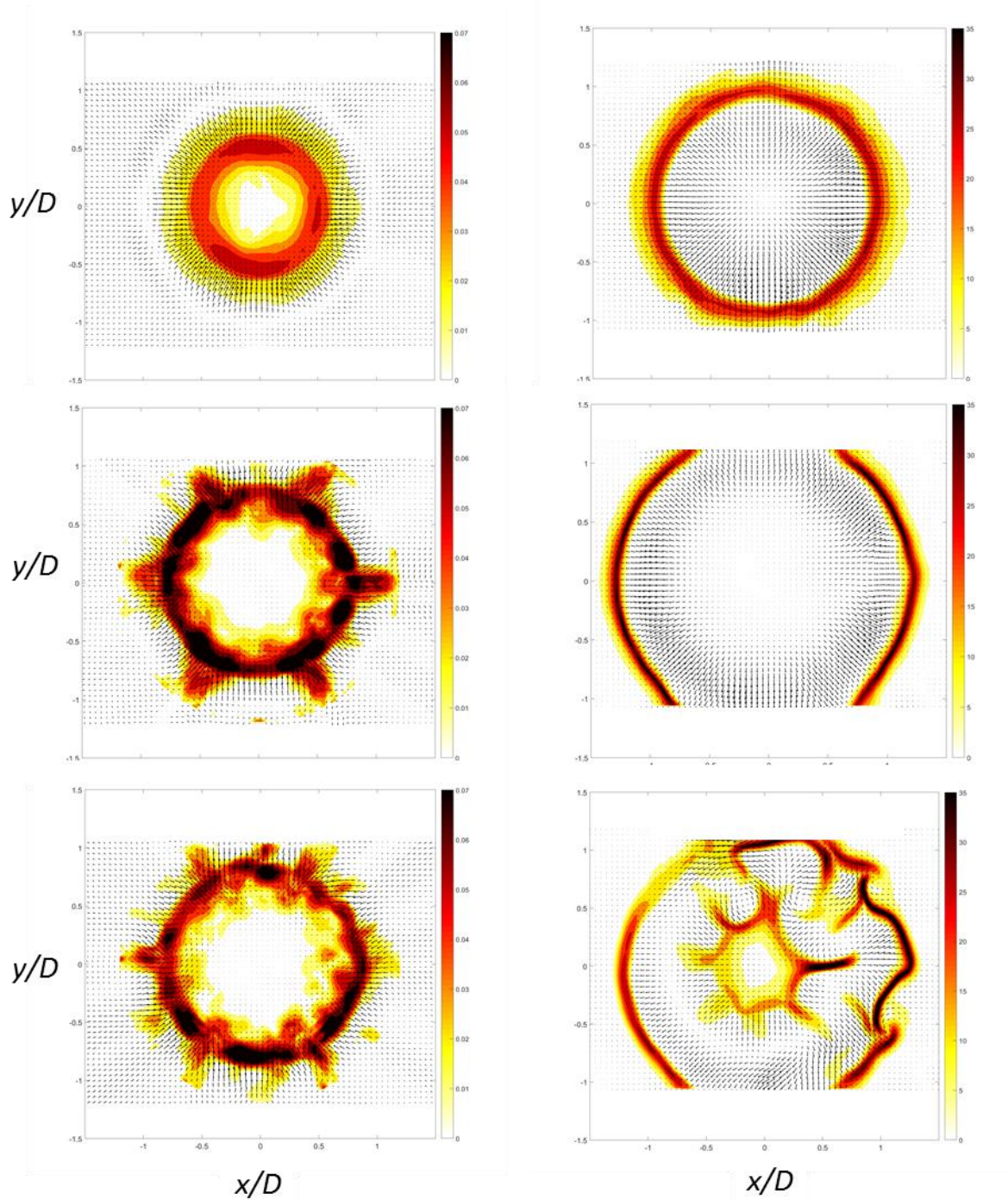

**Figure 6**. Evolution of the fooprint left by the impacting vortex ring in the horizontal *x-y* plane ($z/D$= 0.09), illustrated by the colored FLT-exponent ($T$ = 2.7, $T$=6.1, $T$=9.5). Left: fakir-geometry with hexagonal lattice, right: flat wall.

In the radial plane, interpretation of the FTL-exponent is more difficult as the fluid has a larger component of motion perpendicular to the plane. Near the wall, high FTL-E values indicate the zone of flow separation and/or reattachment at the surface. In the rough wall case, high values occur at the inner edge of the core of the primary ring as revealed by comparison with the vertical plane. This is where fluid is pushed towards the wall. In contrast, for the plane wall high FTL-E values occur at the outer edge of the primary ring where separation takes place and boundary layer fluid is lifted up. Note that in both cases the centre of the primary vortex core remains approximately at *r/D* in the range 0.9-1.0 over the displayed period. As seen in Fig 6, the rough surface forces the early formation of regular disturbance on the primary ring at *T*=6.1 into a pattern similar to a ship's wheel with *N*=6 lobes extending outwards and forming radial wall-jets at their tips. In comparison, for the plane surface the flow expands in radial direction with almost axisymmetric shape until first instabilities appear later at *T*=9.5 in the inner region. At the front of the lobes in Fig. 6 at *r/D*=1, the velocity field shows pairs of counter-rotating vortices which induce the radial jet flows at the tip of the lobes. The obvious hexagonal arrangement allows to apply the decomposition method given in eq. (1) to highlight the arrangement of these vortices. The mode number *N*=6 pattern in azimuthal direction is the most obvious one and the disturbance flow field from eq. (2) then contains less than 10% of the total energy of the flow. The resulting dominant flow pattern is shown in Figure 7 for the two different orientations of the hexagonal lattice as shown in Fig 2A and 2B.

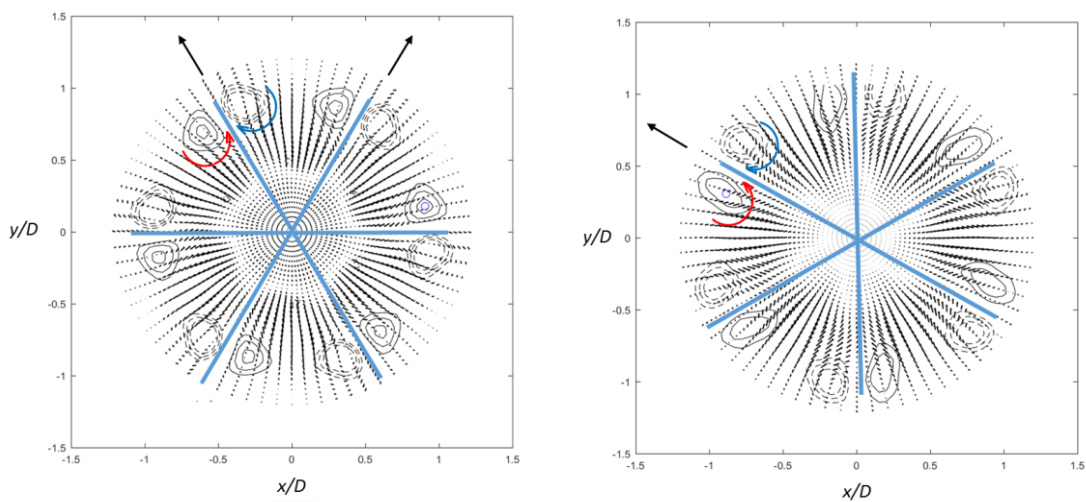

**Figure 7.** Regular arrangement of secondary structures in form of vortex pairs for the hexagonal lattice in original position Fig 2A (left) and for the rotated one in Fig 2B (right) at *T*=5. The blue lines indicate the orientations in the hexagonal lattice with preferred "pathways" in Fig 2A and Fig 2B. The black arrows pointing outwards indicate the wall-jets. Data are obtained after applying eq. 1 to the original data shown in Fig 6. Contour lines indicate wall-normal component of vorticity in regions of concentrated vorticity (dashed lines: negative values; solid lines: positive value; increment is in steps of max/10 from 70% to 100% of max).

The locations of the vortex pairs are displayed by iso-contours of the wall-normal component of the vorticity above the tips of the bed. These regions represent the footprint of a pair of secondary vortices at each lobe which resemble those documented on a free travelling vortex ring as a consequence of non-linear late-stage transition of the primary vortex ring with undulations at a dominant wavenumber *N*=6 (see figure 11 in Ponitz et al. 2016).

Fig. 6 and 7 demonstrate that the lobes are aligned along the orientations of "free" pathways between neighbouring rows of posts, which are the consequence of the hexagonal lattice and thus form directions of higher permeability in the layer. The counter-rotating vortex pairs at the front of the lobes induce at their outer side radial wall-jets directed outwards. Rotating the fakir geometry around the centre of impact also rotates the lobes and the jets' orientation accordingly, see Fig. 7.

Experiments with a vortex ring generated at lower piston velocity ($Re_\Gamma$ = 800) show in general a similar behaviour. In addition, the increase of the height of the posts form 3d to a maximum of 10d doesn't change the general behaviour to form these regular arrangement of the lobes. What matters is the arrangement of the posts. This can be seen in Fig. 8 by comparing the near-wall footprint of the vortex impact to a surface with a random arrangement of the posts at the same number density (same average porosity of $\epsilon = 0.94$).

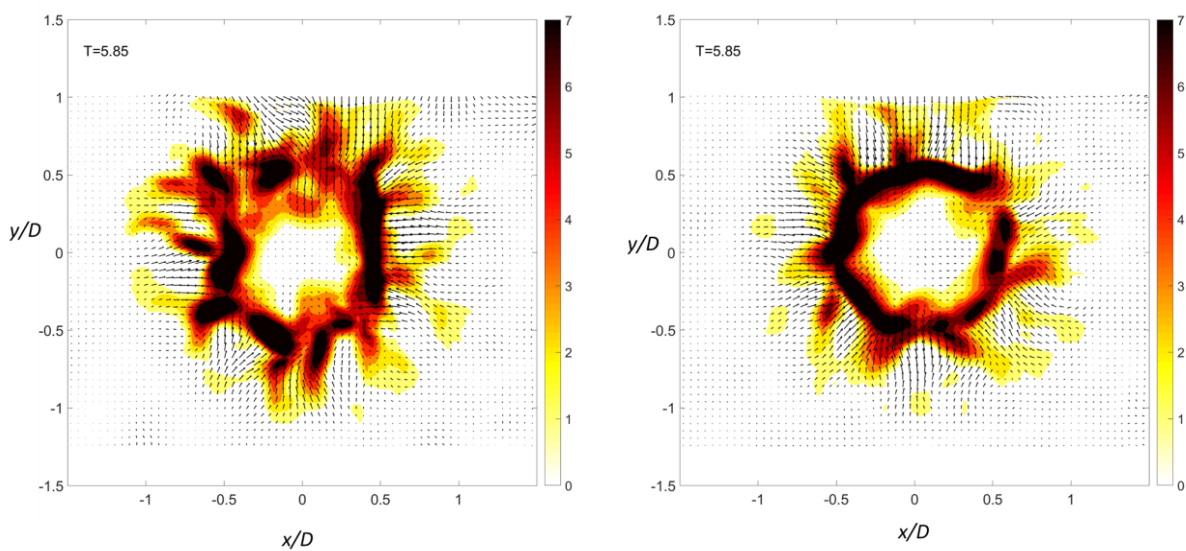

**Figure 8.** Fooprint left by the impacting vortex ring in the horizontal *x-y* plane (*z/D*= 0.09), illustrated by the colored FLT-exponent (*T* = 5.85). Left: fakir-geometry with random lattice of posts, right: the same random lattice rotated about π/2.

Unlike the hexagonal lattice, the formation of secondary structures in the radial plane is not correlated to the random arrangement of the posts as we see from the comparison with the rotated wall in Fig 8. As a consequence, there is seemingly no lock-on with the instabilities of the vortex ring.

## 4. Discussion

In summary, the observed flow pattern after impact of the vortex ring is strongly correlated with the hexagonal lattice structure and its orientation. The instabilities of the vortex ring lock-on with the preferential "pathways" in the lattice, which is designed herein to enforce mode number *N*=6 in azimuthal direction. For the given vortex ring properties, the mode *N*=6 is the one which dominates in the later stage of the transition phase according to the theory of Saffmann (1978). With impact on the hexagonal lattice of posts, the non-linear process of generation of undulations in the primary ring is triggered much earlier. This generates at the outer edge of the primary ring a regular arrangement of radial wall-jets. When the lattice is irregular, no regular pattern of secondary vortices is recognizable.

Note that in the experiments with the rough wall, the results in the vertical flow plane do not show the typical interaction of the second vortex ring being wrapped around the primary one. Rather, we observe the radial expansion of the secondary vortex ring just after formation, while the core of both rings remain approximately in the same plane. It is known that, when a vortex approaches a no-slip wall, it induces a layer of vorticity of opposite sign which, under certain circumstances, may roll into new vortices. Herein, the roll-up of the boundary layer is not as coherent as in the flat wall case because radial disturbances are built up quickly in form of the lobes. As the secondary vortex ring is formed next to the primary one and remains in the same plane, a strong vertical rebound of the fluid between both regions of opposite sign vorticity is induced, much earlier than in the case of the flat wall.

For plane walls, both numerical simulation (Orlandi and Verzicco 1993) and experimental studies (Lim et al. 1991) confirmed that the growth rate of secondary vortex instabilities was much larger than that of primary vortex. These seemed to indicate that the secondary vortex instabilities triggered the breakdown process. Furthermore, it was argued that the transition happens after the secondary vortex ring convected into the interior of the primary vortex ring. Indeed, the formation of instabilities in the flat wall experiment shows that the instabilities occur in the inner of the primary ring as evidenced in Fig 6. However, such a behaviour is not observed for the rough wall because the secondary vortex moves radially away from the primary one. This let us conclude that the observed lock-on process herein is related to the instabilities in the primary vortex ring, which are triggered by the impact of the structured wall and undergo rapid non-linear growth with formation of the lobes.

## 5. Conclusion

We have investigated the effect of impact of a vortex ring on a wall with tailored arrangements of slender posts, resembling a fakir geometry. For a fixed, low Reynolds number and constant height of the posts with *h/D*=0.068 the flow in the horizontal plane near the interface is investigated with TR-PIV. In addition, measurements in the vertical centre-plane were done to characterize the flow evolution in the wall-normal symmetry plane. These results were compared with experiments of the impact on a plane smooth wall. For the rough wall a secondary vortex is generated that is radially expanding and moving away from the primary one, causing a strong fluid rebound induced by the opposite sign vorticity next to each other. The primary vortex undergoes early transition triggered by the impact and radial expansion is smaller than in case of the plain wall. A consistent evolution of azimuthal instability mode *N*=6 is achieved with the hexagonal arrangement while no clear dominance of modes evolves using a random arrangement of the posts. Differently also from the random arrangement, the phase of azimuthal mode *N*=6 is enforced by the design of the hexagonal lattice, having "preferential" pathways. Thus, the phase of the mode 6 arrangement is always locked with the hexagonal lattice. As evidenced in the experiments, if the fakir geometry is rotated, the phase rotates accordingly. This behaviour suggests that tailored lattices made from roughness elements as used in form of slender posts can be used for manipulation of instability modes and their phase.

As shown herein, a flow control device can be thought of which forms wall-jets in a certain directional pattern, e.g with a hexagonal lattice tailored to the instability modes of the vortex ring to enforce six radial wall-jets at the outer tip of the lobes. Furthermore, such surfaces could also be useful for the near-wall control of instabilities in turbulent boundary layer flows. It was shown in previous studies that the arrangement of flexible micro-rods in rows aligned with the mean flow can lead to increased coherence in spanwise direction which is an indication to drag-reducing behaviour (Brücker 2011). Mentioned therein, the consequence of the specific arrangement of the micro-hairs in streamwise columns led to a reduced spanwise meandering and stabilization of the streamwise velocity streaks. Therefore the surface with the streamwise aligned lattice is promoting varicose waves and inhibiting sinusoidal waves. Streak stabilization is known to be a major contributor to turbulent drag reduction.

It was recently argued by Rosti et al. (2018) with respect to turbulent drag reduction that porous materials with high wall-normal and low wall-parallel permeabilities, e.g., a carpet of wall-normal rods, are characterized by an increased turbulence isotropy in the near-to-the-interface region with a consequent disruption of the streamwise coherence probably due to the emergence of alternating spanwise correlated structures leading to an increased drag. However, their study does not include the effects studied herein, which is considering large directional anisotropy of wall-parallel permeability in the layer. Therefore, the conclusion of the study of Rosti et al. (2018) is not complete. Rather, one needs to take into account the impact of such arrangements as those of the fakir-type studied herein on instabilities in the flow. As we could show here, instabilities can be manipulated towards selection of dominant modes and lock-on their phase by specific arrangement of the elements. Such surfaces with posts can be easily modified in their arrangement, geometry, scale and spacing using modern additive printing technology, therefore it is a charming alternative of a passive flow control technique compared to porous layers which are difficult to control in their individual local structure and anisotropy of permeability.

# Acknowledgment

This work was sponsored by BAE System and the Royal Academy of Engineering, both jointly funding the position of Professor Christoph Bruecker, which is gratefully acknowledged herein. The position of

MSc Qianhui Li was supported by the German Research Foundation in the grant DFG 1494/31-2 which is also gratefully acknowledged. We thank Prof. Uwe Schnakenberg from IWE1, RWTH Aachen, Germany, for support in building the transparent fakir geometry. Finally, we thank Dr. Zhengzhong Sun for his support in the finite-time Lyapunov exponent processing.

# References

Adhikari, D. and Lim, T. T., "The impact of a vortex ring on a porous screen," Fluid Dyn. Res. **41** 051404, (2009).

Brücker, C. H., "Interaction of flexible surface hairs with near-wall turbulence," J. Phys. Condens. Matter, 23(18), (2011).

Chu, C. C., Wang, C. T., and Chang, C. C., "A vortex ring impinging on a solid plane surface—Vortex structure and surface force," Physics of Fluids 7, 1391, (1995).

Fransson, J. H. M, Talamelli, A., Brandt, L., and Cossu, C., "Delaying Transition to Turbulence by a Passive Mechanism," Phys. Rev. Lett. **96**, 064501, (2006).

Jiménez, J., "On the structure and control of near wall turbulence," Physics of Fluids 6, 944 (1994).

Jiménez, J. and Pinelli, A., "The autonomous cycle of near-wall turbulence," J. Fluid Mech. 389, pp. 335-359, (1999).

Lim, T.T., Nickels, T. B. and Chong, M. S., "A note on the cause of rebound in the head-on collision of a vortex ring with a wall," Exp. in Fluids 12(1-2), pp. 41–48, (1991).

Orlandi and R. Verzicco, R., "Vortex rings impinging on walls: Axisymmetric and three-dimensional simulations," J. Fluid Mech. **256**, 615 (1993).

Pointz, B., Sastuba, M. and Bruecker, C. H., "4D visualization study of a vortex ring life cycle using modal analysis," J. Visualization. 19:237:259. (2016)

Rosti, M., Brandt, L. and Pinelli, A., "Turbulent channel flow over an anisotropic porous wall - Drag increase and reduction," arXiv:1802.00477 [physics.flu-dyn] , (2018)

Walker, J. D. A., Smith, C. R., Cerra, A. W. and Doligalski, T. L., " The impact of a vortex ring on a wall," J. Fluid Mech. **181**, 99-140 (1987).

Saffman, P.G., "The number of waves on unstable vortex rings," J. Fluid Mech. 84:625-639. (1978).

Sun, Z. and Bruecker, C. H., "Investigation of the vortex ring transition using Scanning Tomo-PIV," Exp. in Fluids, 58, 36, (2017).

Xu, Y., Wang, J., "Recent development of vortex ring impinging onto the wall," Science China Technological Sciences 56(10), pp 2447–2455, (2013).